\documentclass[twocolumn,preprintnumbers,amsmath,amssymb,aps,pra]{revtex4-1}

\usepackage[utf8]{inputenc}
\usepackage{graphicx}

\usepackage{color}

\newcommand*{\mat}[1]{\mathbf{#1}}
\newcommand*{\vecr}{\mat{r}}
\newcommand*{\rref}{\vecr_{\text{ref}}}

\newcommand*{\integ}[1]{\int\!\!\!\:\ud{#1}\:}
\newcommand*{\iinteg}[2]{\integ{#1}\!\!\!\integ{#2}}

\newcommand*{\crea}[1]{\hat{#1}^{\dagger}}
\newcommand*{\anni}[1]{\hat{#1}^{\vphantom{\dagger}}}

\newcommand*{\leftnabla}{%
  \raisebox{1.5ex}{\makebox[-0.5pt][l]{$\scriptstyle \leftharpoonup$}}\nabla}
\newcommand*{\rightnabla}{%
  \raisebox{1.5ex}{\makebox[-0.3pt][l]{$\scriptstyle \rightharpoonup$}}\nabla}

\newcommand{\abraket}[2]{\left\langle{#1}\middle|{#2}\right\rangle}

\newcommand{\brakket}[3]{\langle{#1}|{#2}|{#3}\rangle}

\newcommand*{\abs}[1]{\lvert#1\rvert}
\newcommand*{\du}{\partial}
\newcommand*{\e}{\textrm{e}}

\newcommand*{\half}{\frac{1}{2}}
\newcommand*{\ud}{\mathrm{d}}

\begin{document}

\title{Towards non-local density functionals by explicit modeling of the exchange-correlation hole in inhomogeneous systems}
\author{K.J.H. Giesbertz}
\author{R. van Leeuwen}
\affiliation{Department of Physics, Nanoscience Center, University of Jyväskylä, P.O. Box 35, 40014 Jyväskylä, Survontie 9, Jyväskylä, Finland}
\author{U. von Barth}
\affiliation{Department of Mathematical Physics, Institute of Physics, Lund University, Sölvegatan 14A, Lund S-22362, Sweden}

\date{\today}

\begin{abstract}
We put forward new approach for the development of a non-local density functional by a direct modeling of the shape of exchange-correlation (xc) hole in inhomogeneous systems. The functional is aimed at giving an accurate xc-energy and an accurate corresponding xc-potential even in difficult near-degeneracy situations such as molecular bond breaking. In particular we demand that: (1) the xc hole properly contains $-1$ electron, (2) the xc-potential has the asymptotic $-1/r$ behavior outside finite systems and (3) the xc-potential has the correct step structure related to the derivative discontinuities of the xc-energy functional. None of the currently existing functionals satisfies all these requirements. These demands are achieved by screening the exchange hole in such a way that the pair-correlation function is symmetric and satisfies the sum-rule. These two features immediately imply (1) and (2) while the explicit dependence of the exchange hole on the Kohn--Sham orbitals implies (3). Preliminary calculations show an improved physical description of the dissociating hydrogen molecule. Though the total energy is still far from perfect, the binding curve from our non-local density functional provides a significant improvement over the local density approximation.
\end{abstract}

\maketitle

\section{Introduction}
\label{sec:introduction}

The local density approximation (LDA) is the simplest functional in density functional theory (DFT) and has been around since the advent of DFT~\cite{HohenbergKohn1964, KohnSham1965b}. Although the LDA has only a local dependence on the density, it has been tremendously successful in describing the ground-state properties of solids, surfaces and large molecules. Its shortcomings have been obvious from the beginning and an enormous effort has gone into the search for better approximations. This task has proved to be exceedingly difficult. One should, however, not be surprised or disappointed. Density functional theory provides a way to reduce the full, interacting many-body problem to a simple non-interacting one. Therefore, an accurate density functional for the total energy would provide a surprisingly simple way to solve the complicated many-body problem --- at least as far as static properties are concerned. Nevertheless, in the past decades considerable progress has been made. With the advent of the generalized gradient approximations (GGAs)~\cite{Becke1988, Perdew1991}, bond lengths and atomization energies were greatly improved as compared to those of the LDA. Unfortunately, the exchange-correlation (xc) potentials of the GGAs have several undesirable features. In particular, they decay too fast outside finite systems~\cite{LeeZhou1991, EngelChevaryMacdonald1992}, unlike a correct $-1/r$ decay. Consequently, neither the LDA nor the GGAs produce proper Rydberg levels~\cite{LeeuwenBaerends1994} and ionization potentials are too low.

An important class of extensions to the GGAs came from the observation that exchange energies are much larger than correlation energies. This suggests that exchange should be treated exactly, while using a GGA only for the correlation energy. Full inclusion of ``exact exchange'' does however not work well in practice, since there is a large cancellation of errors between the LDA or the GGA versions of the exchange and correlation energies. On the other hand, using only a portion of exact exchange combined with a GGA does yield quite accurate bond energies in molecules~\cite{Becke1993}. An important advantage of ``exact exchange'' is that it provides some necessary improvements of the xc potential. For instance, the inclusion of ``exact exchange'' gives a stepped structure in the xc potential~\cite{LeeuwenGritsenkoBaerends1995, LeeuwenGritsenkoBaerends1996} thus improving the description of the atomic shells as well as enabling a neutral dissociation of heteronuclear molecules~\cite{PerdewParrLevyBalduz1982, GritsenkoBaerends1996, BaerendsGritsenko1997}. It also gives an xc-potential with the proper asymptotic ($-1/r$) behavior which gives rise to Rydberg levels and a highest occupied KS eigenvalue in better agreement with the negative of the ionization potential. Usually, however, only a fraction of full exchange is incorporated in the so called hybrid functionals meaning that the desirable features mentioned above are only partially obtained. It seems that only functionals with a massive amount of fitting parameters are able to handle 100\% ``exact exchange''~\cite{Gill2001, ZhaoTruhlar2008}, though they will always suffer from weak singularities in the response functions of metals. We mention here that inclusion of ``exact exchange'' is not mandatory in order to have good properties of the xc potential. The proper step structure as well as the correct asymptotics away from finite systems can also be obtained by modeling the potential directly~\cite{LeeuwenBaerends1994, GritsenkoLeeuwenLenthe1995}. Such model potentials can indeed provide good response properties~\cite{GritsenkoSchipperBaerends1999, SchipperGritsenkoGisbergen2000} but they cannot easily be written as the functional derivative of some accurate functional for the xc energy. As a result, their implementations have been limited. Further improvements to the energy functional are also sought by adding the kinetic-energy density to the functional arguments~\cite{GhoshParr1986, TaoPerdewStaroverov2003} and including parts from many-body perturbation theory~\cite{GorlingLevy1994, Leeuwen1996, GattiOlevanoReining2007, RomanielloSangalliBerger2009, HellgrenBarth2010}.

In this paper we would like to consider an alternative approach by looking directly at electron correlations in real space. In a correlated system we can consider the conditional probability of finding an electron at some point in space, when the position of another reference electron is given. The difference between this function and the unconditional probability (which is simply the density) is defined to be the xc hole and the knowledge of this function~\footnote{To be precise we need to know xc hole for different strengths of the electron interaction.} is sufficient to calculate the xc-energy. 

The effect of exchange and correlation is to dig a hole in the density around each electron, so as to remove one electron in that region. We can say that the task of a good xc functional is to provide an accurate description of the xc hole. The LDA and GGA assume that this hole has a spherical shape and an extent given by the Wigner--Seitz radius ($4\pi r_s^3 n = 3$, $n$ being the electron density) at the reference electron. Unfortunately for the LDA, xc holes are definitely not spherical~\cite{PhD-Buijse1991, BuijseBaerends2002}. Although hybrid functionals lift this restriction, their blunt use of ``exact exchange'' actually worsens the description of the xc hole for stretched bonds compared to the LDA and GGA functionals.

A more sophisticated class of functionals which aims to build a non-spherical model of the xc hole is the so-called weighted density approximations (WDA)~\cite{AlonsoGirifalco1978, GunnarssonJonsonLundqvist1979, SaddTeter1996, RushtonTozerClark2002}. These functionals avoid the spherical xc hole by digging a hole out of the real density rather than in the density at the position of the reference electron. A nice feature of the WDA is that the asymptotic behavior of its xc potential has a Coulombic asymptotic decay instead of an exponential behavior as in the LDA and the GGA. An important symmetry of the pair-density ($\Gamma(\vecr',\vecr) = \Gamma(\vecr,\vecr')$) is, however, broken. This causes an additional factor $1/2$ in the asymptotic decay of the xc potential, so that it decays too fast (as $-1/(2r)$)~\cite{OssiciniBertoni1985}.

Here, we will advocate an approach in a similar spirit as the weighted density approximation~\cite{Barth2004}. However, we will take care not to destroy the symmetry of the pair-density and therefore, the xc potential will automatically have the correct asymptotic $-1/r$ behavior. Furthermore, important information on the physics of the xc hole is provided by the dissociation of molecules. In particular, a proper localization of the xc hole around the reference electron for a dissociated molecule is a challenging task. The failure of current approximations to achieve this is reflected in their consistent inability to properly describe the breaking of chemical bonds. The most important aim of our new functional will therefore be a bold one: the new functional has to be able to describe molecular dissociation.

The paper is outlined as follows. First (Sec.~\ref{sec:motivation}) we will give a more detailed discussion on the background to motivate the construction of the screened exchange (SX) functional. The actual construction of the SX functional is discussed in section~\ref{sec:SXmodel}. In section~\ref{sec:results} we show preliminary results and finally conclude in section~\ref{sec:conclusion}.

\section{Motivation}
\label{sec:motivation}

\subsection{Symmetry and asymptotics}
\label{sec:SymAndAsymp}

We will start from an exact expression~\cite{Almbladh1972, LangrethPerdew1975, GunnarssonLundqvist1976} for the exchange-correlation energy
\begin{align}\label{eq:xcEnergy}
E_{\text{xc}} 
= \half \iinteg{\vecr}{\vecr'}n(\vecr)\frac{n(\vecr')\bigl(\bar{g}(\vecr,\vecr') - 1\bigr)}{\abs{\vecr-\vecr'}},
\end{align}
where $n(\vecr)$ is the density. This expression gives the exact xc contribution to the interaction energy of the system, provided $\bar{g}$ is the exact pair-correlation function $g$ of the system, defined as
\begin{align*}
g(\vecr,\vecr') \equiv \frac{\Gamma(\vecr,\vecr')}{n(\vecr)n(\vecr')}.
\end{align*}
Here the diagonal of the two-body reduced density matrix is defined as
\begin{align*}
\Gamma(\vecr,\vecr') &\equiv \sum_{\sigma,\sigma'}\Gamma(\vecr\sigma,\vecr'\sigma') \notag \\
&\equiv \sum_{\sigma,\sigma'}
\brakket{\Psi}{\crea{\psi}(\vecr\sigma)\crea{\psi}(\vecr'\sigma')
\anni{\psi}(\vecr'\sigma')\anni{\psi}(\vecr\sigma)}{\Psi},
\end{align*}
where $\crea{\psi}(\vecr\sigma)$ and $\anni{\psi}(\vecr\sigma)$ are the usual field operators. The xc energy, however, also contains a correlation contribution to the kinetic energy which is most conveniently included by integrating the interaction energy with respect to the strength $\lambda$ of the Coulomb interaction --- while keeping the density fixed at the fully interaction one ($\lambda=1$)~\cite{Almbladh1972, LangrethPerdew1975, GunnarssonLundqvist1976, Barth2004}, i.e.
\begin{align*}
\bar{g}(\vecr,\vecr') \equiv \int_0^1\!\!\!\ud\lambda\, g_{\lambda}(\vecr,\vecr').
\end{align*}
The physical picture of representing the xc energy in this manner is that an electron does not interact with the full density, but depending on its position, $\vecr$, it sees an effective density $n(\vecr')\bar{g}(\vecr,\vecr')$ with $N - 1$ electrons. Therefore, the part
\begin{align}\label{eq:xcHoleDef}
\bar{\rho}_{\text{xc}}(\vecr'|\vecr) \equiv n(\vecr')\bigl(\bar{g}(\vecr,\vecr') - 1\bigr) 
\end{align}
in the xc energy~\eqref{eq:xcEnergy} exactly describes the density of minus one electron, a hole, which is reflected in the sum-rule of the xc hole
\begin{align}\label{eq:xc holeSumRule}
\integ{\vecr'} \bar{\rho}_{\text{xc}}(\vecr'|\vecr)
= \integ{\vecr'} n(\vecr')\bigl(\bar{g}(\vecr,\vecr') - 1\bigr) = -1.
\end{align}
The local density approximation (LDA) proceeds by using the xc hole from the homogeneous electron gas evaluated for the density at the position of the reference electron, so the pair-correlation function is approximated as $\bar{g}(\vecr,\vecr') \approx \bar{g}_h\bigl(\abs{\vecr-\vecr'};n(\vecr)\bigr)$. Furthermore, if the distance $\abs{\vecr-\vecr'}$ is large the pair-correlation function $\bar{g}$ only differs slightly from one, so it is quite reasonable to replace $n(\vecr')$ by $n(\vecr)$ in the xc-energy~\eqref{eq:xcEnergy}. Combining both approximations, we obtain the expression for the xc energy of the LDA
\begin{align}\label{eq:LDAenergy}
E_{\text{xc}}^{\text{LDA}}
&= \half \iinteg{\vecr}{\vecr'}n(\vecr)
\frac{n(\vecr)\bigl\{\bar{g}_h(\abs{\vecr-\vecr'};n(\vecr)\bigr) - 1\bigr\}}{\abs{\vecr-\vecr'}} \notag \\
&= \half \iinteg{\vecr}{\vecr'}n(\vecr)
\frac{n(\vecr)\bigl\{\bar{g}_h\bigl(\abs{\vecr'};n(\vecr)\bigr) - 1\bigr\}}{\abs{\vecr'}} \notag \\
&= \integ{\vecr}n(\vecr)\varepsilon_{\text{xc}}\bigl(n(\vecr)\bigr),
\end{align}
where the function $\varepsilon_{\text{xc}}(n)$ is just the exchange-correlation energy per electron of the homogeneous electron gas. This expression for the xc energy of the LDA shows explicitly that its hole is approximated by a spherical one. As mentioned above xc holes are not spherical~\cite{PhD-Buijse1991, BuijseBaerends2002}. The original weighted density approximation (WDA)~\cite{AlonsoGirifalco1978, GunnarssonJonsonLundqvist1979} improves on the shape of the xc hole by not replacing $n(\vecr')$ by $n(\vecr)$ in the xc-energy~\eqref{eq:xcEnergy}. Since the sum-rule~\eqref{eq:xc holeSumRule} for the xc hole is no longer trivially satisfied, an effective density, $\bar{n}(\vecr)$, is used as input into the pair-correlation function instead of the density at the reference position. The xc energy in the WDA is therefore
\begin{align}\label{eq:WDAenergy}
E_{\text{xc}}^{\text{WDA}}
= \half \iinteg{\vecr}{\vecr'}n(\vecr)
\frac{n(\vecr')\bigl\{\bar{g}_h(\abs{\vecr-\vecr'};\bar{n}(\vecr)\bigr) - 1\bigr\}}{\abs{\vecr-\vecr'}},
\end{align}
where the effective density $\bar{n}(\vecr)$ should be found by satisfying the sum-rule for the xc hole~\eqref{eq:xc holeSumRule} at every point~$\vecr$
\begin{align*}
\integ{\vecr'} n(\vecr')\bigl\{\bar{g}_h(\abs{\vecr-\vecr'};\bar{n}(\vecr)\bigr) - 1\bigr\} = -1.
\end{align*}
Unfortunately, it was found that the pair-correlation functional of the homogeneous electron gas is not a good approximation to the pair-correlation function of inhomogeneous systems as molecules and surfaces. Using a more localized function for $\bar{g} - 1$, the results were significantly improved~\cite{GunnarssonJones1980, SaddTeter1996, RushtonTozerClark2002}.

From the expression for the WDA xc energy~\eqref{eq:WDAenergy} we immediately notice that the integral kernel is still asymmetric in the coordinates $\vecr$ and $\vecr'$ and therefore breaks the important symmetry $\bar{g}(\vecr,\vecr') = \bar{g}(\vecr',\vecr)$. This is not so important for the value of the xc energy. But for the corresponding xc potential, we obtain
\begin{multline*}
v_{\text{xc}}^{\text{WDA}}(\vecr)
= \half \integ{\vecr'}
\frac{n(\vecr')\bigl\{\bar{g}_h(\abs{\vecr-\vecr'};\bar{n}(\vecr)\bigr) - 1\bigr\}}{\abs{\vecr-\vecr'}}  \\
{} + \half \integ{\vecr'}
\frac{n(\vecr')\bigl\{\bar{g}_h(\abs{\vecr-\vecr'};\bar{n}(\vecr')\bigr) - 1\bigr\}}{\abs{\vecr-\vecr'}}  \\
{} + \half \iinteg{\vecr'}{\vecr''}n(\vecr')\frac{n(\vecr'')}{\abs{\vecr'-\vecr''}} \\
{} \times \frac{\delta\bar{g}_h(\abs{\vecr'-\vecr''};\bar{n}\bigr)}{\delta \bar{n}(\vecr'')}
\frac{\delta\bar{n}(\vecr'')}{\delta n(\vecr)}.
\end{multline*}
The first term actually decays as $-1/(2r)$ as is obvious from the sum-rule. The long-range behavior of the other two terms is not so obvious, but in practice they turn out to decay exponentially~\cite{SaddTeter1996, RushtonTozerClark2002}. Therefore, the xc potential decays too slowly (as $-1/(2r)$) compared to the proper $-1/r$ decay~\cite{GunnarssonJonsonLundqvist1979}. The incorrect long-range behavior of the potential is expected to have a significant effect on properties which depend strongly on a proper xc potential such as the ionization energy and Rydberg excitations~\cite{AlmbladhBarth1985, LeeuwenGritsenkoBaerends1995, SchipperGritsenkoGisbergen2000}. This failure can be attributed to the broken symmetry of the pair-correlation function. Consequently, one of the requirements of our new functional will be that it should satisfy the proper symmetry of the pair-correlation function, i.e.\ $\bar{g}(\vecr,\vecr') = \bar{g}(\vecr',\vecr)$.

\subsection{Step structure from exchange}
As mentioned in the introduction, the steps in the KS potential are important for a proper description of the atomic shell structure~\cite{LeeuwenGritsenkoBaerends1995, LeeuwenGritsenkoBaerends1996} and also the neutral dissociation of hetero-nuclear molecules~\cite{PerdewParrLevyBalduz1982, GritsenkoBaerends1996}. It has been shown that the necessary steps for the atomic shell structure are already featured by the exchange energy~\cite{LeeuwenGritsenkoBaerends1995}. However, the necessary step in the dissociation of hetero diatomic molecules is a less clear case, since the spin-symmetry has to be broken to provide the necessary localization~\cite{FuksRubioMaitra2011, MakmalKummelKronik2011}.

\begin{figure*}[t]
  \includegraphics[width=\textwidth]{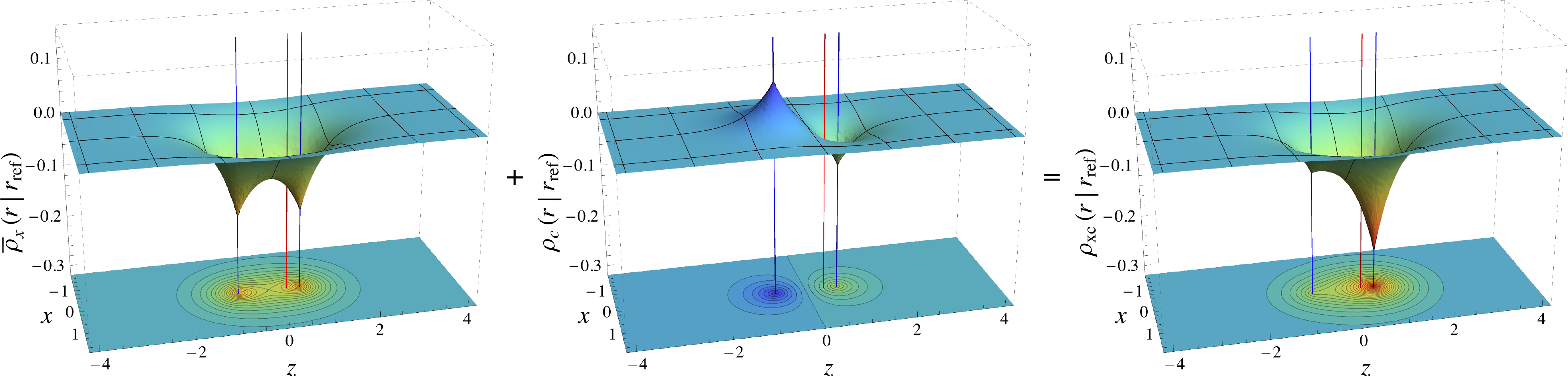} \\
  \caption{(Color online) The different holes of the H$_2$ molecule at $R_{\text{H--H}} = 1.4$ Bohr in atomic units and the reference electron at 0.3 Bohr to the left of the right nucleus along the bond axis ($\rref = (0,0,0.4)$ Bohr). The positions of the nuclei are indicated by the blue lines and the position of the reference electron is indicated by the red line. The left panel shows the exchange hole, $\bar{\rho}_{\text{x}}(\vecr|\rref) = -\abs{\sigma_g(\vecr)}^2$, the middle panel shows the correlation hole, $\rho_{\text{c}}(\vecr|\rref)$, which provides a small correction to have the more localized real hole, $\rho_{\text{xc}}(\vecr|\rref)$.}
  \label{fig:H2holesEqui}
  \includegraphics[width=\textwidth]{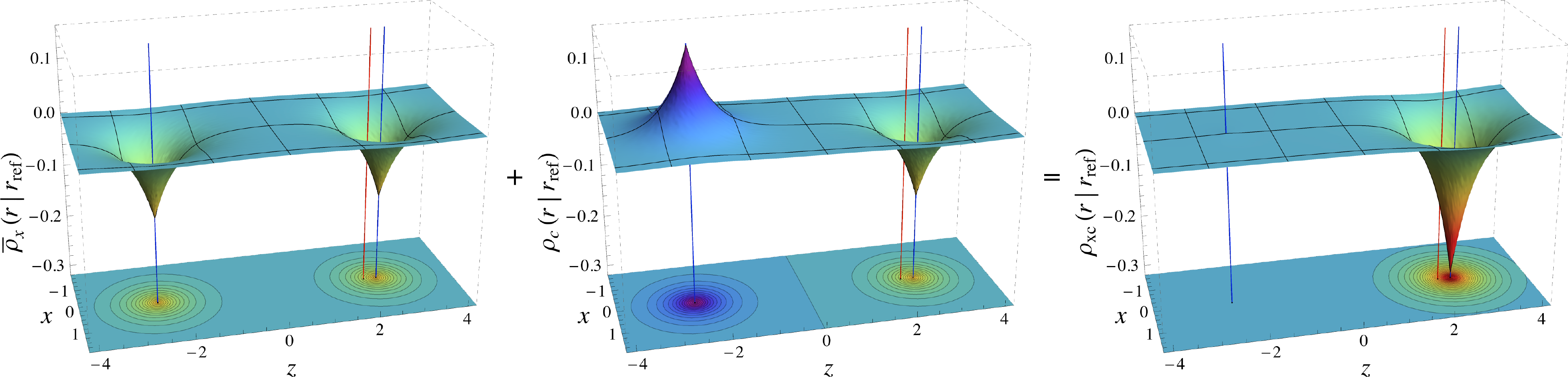}
  \caption{(Color online) Similar to the previous plots, but now for $R_{\text{H--H}} = 5.0$ Bohr. The reference electron is still at 0.3 Bohr to the left of the right nucleus along the bond axis ($\rref = (0,0,2.2)$ Bohr now). The exchange hole, $\bar{\rho}_{\text{x}}(\vecr|\rref)$, remains completely delocalized, so requires an equally large correction from the correlation hole, $\rho_{\text{c}}(\vecr|\rref)$, to obtain the real hole, $\rho_{\text{xc}}(\vecr|\rref)$, localized around the reference electron.}
  \label{fig:H2holes5}
\end{figure*}

Closely related, the exchange energy also has the necessary features for the integer discontinuity~\cite{LeeuwenGritsenkoBaerends1995, GritsenkoBaerends1996}, since the exchange term often changes radically when crossing an integer number of electrons due to the usual idempotency of the KS density matrix. The corresponding hole, the exchange hole, can be expressed in spin-integrated quantities as
\begin{align}\label{eq:xHoleDef}
\bar{\rho}_{\text{x}}(\vecr|\rref) \equiv -\half\frac{\abs{\gamma_s(\vecr,\rref)}^2}{n(\rref)},
\end{align}
where the spin-integrated KS density matrix is defined as
\begin{align*}
\gamma_s(\vecr,\vecr') &\equiv \sum_{\sigma}\gamma(\vecr\sigma,\vecr'\sigma) \notag \\
&\equiv \sum_{\sigma}\brakket{\Psi_s}{\crea{\psi}(\vecr'\sigma')\anni{\psi}(\vecr\sigma)}{\Psi_s},
\end{align*}
with $\Psi_s$ as the KS wavefunction. The KS density matrix can alternatively be expressed directly in terms of the KS orbitals $\phi_k(\vecr)$ as
\begin{align}\label{eq:KS1RDM}
\gamma_s(\vecr,\vecr') = \sum_kn_k\phi_k(\vecr)\phi^*_k(\vecr'),
\end{align}
where $n_k$ are the occupation numbers, being simply 0 or 2 in the non-degenerate case. The exchange hole has the convenient property that it already satisfies the hole xc hole sum-rule~\eqref{eq:xc holeSumRule}.

Since exchange already satisfies a number of important properties, it is often used as a starting point to model the full exchange-correlation effects. Traditionally, one adds a correction term, correlation, defined as the \emph{difference} between the exact exchange-correlation quantities and the ones with one exchange taken into account. For example the correlation hole is simply defined as
\begin{align*}
\bar{\rho}_{\text{c}}(\vecr|\rref) \equiv \bar{\rho}_{\text{xc}}(\vecr|\rref) - \bar{\rho}_{\text{x}}(\vecr|\rref).
\end{align*}
Although this approach has some appeal, it is inconvenient in practice, especially in bond-breaking situations. We show that explicitly in the next section.

\subsection{Bond breaking}

\begin{figure*}[t]
  \includegraphics[width=\textwidth]{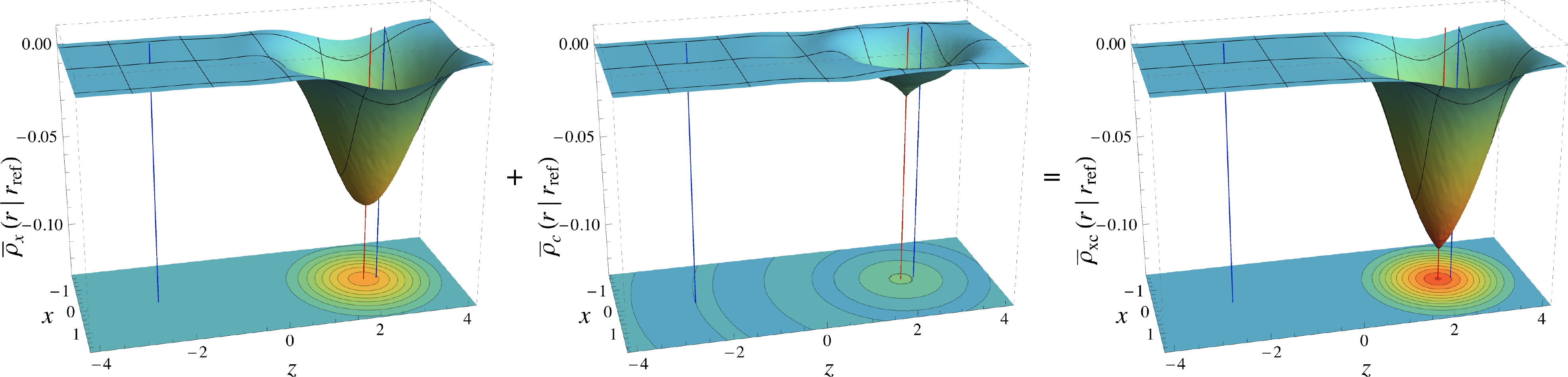}
  \caption{(Color online) LDA holes for the H$_2$ molecule at $R_{\text{H--H}} = 5.0$ Bohr in atomic units. The reference electron is again at 0.3 Bohr to the left of the right nucleus along the bond axis ($\rref = (0,0,2.2)$. Both the LDA x hole and c hole are localized, so do not resemble the exact ones. However, their sum, $\rho_{\text{xc}}(\vecr|\rref)$, has much better resemblance to the full xc hole.}
  \label{fig:LDAholes5}
\end{figure*}

In Fig.~\ref{fig:H2holesEqui} we have plotted the different holes, $\bar{\rho}_{\text{x}}(\vecr|\rref)$, $\rho_{\text{c}}(\vecr|\rref)$ and $\rho_{\text{xc}}(\vecr|\rref)$ for the H$_2$ molecule at equilibrium distance $R_{\text{H--H}} = R_{\text{e}} = 1.4$ Bohr. Note that the quantities with correlation are at full coupling strength ($\lambda = 1$) and not the integrated ones. Ideally we would like to have shown the integrated ones, but obtaining them is a rather difficult task. Since the effect of the kinetic energy is not very large on the total energy~\cite{PhD-Leeuwen1994}, we expect that the holes do not differ too much as well. The reference electron is fixed at 0.3 Bohr to the left from the right nucleus. The holes were calculated from full configurations interaction (CI) results using the 1s, 2s, 3s, 2p and 3d hydrogen wavefunctions on each atom as a basis set
\footnote{The 3p orbitals caused linear dependency problems at $R_{\text{H--H}} = R_{\text{e}} = 1.4$ Bohr, so they could not be included.}. 
The exchange hole (x hole) can be worked out to be
\begin{align*}
\bar{\rho}_{\text{x}}(\vecr|\rref) = -2\frac{\abs{\sigma_g(\vecr)}^2\abs{\sigma_g(\rref)}^2}{n(\rref)} 
= -\abs{\sigma_g(\vecr)}^2,
\end{align*}
thus the exchange hole is actually independent of the position of the reference electron. However, the real hole is deeper around the reference electron and therefore, depending on the position of the reference electron, the correlation hole (c hole) has to add and remove an equal amount of the hole to deepen it around the reference electron. Although the xc hole is more localized around the reference position, it still shows the two-peak structure of the x hole.

The localization effect becomes more prominent if we consider the hydrogen molecule at an elongated bond distance of $R_{\text{H--H}} = 5.0$ Bohr (Fig.~\ref{fig:H2holes5}). Again, the x hole is independent of the position of the reference electron and is completely delocalized over the molecule. However, the real hole is completely localized around the reference electron, which is again located at 0.3 Bohr to the left from the right nucleus. Therefore, the c hole has to completely remove the hole from the left side of the molecule at put it back at the right side such that it integrates still to $-1$ electron. The c hole can not be regarded as a ``small'' correction to the x hole anymore, since it is actually equal in magnitude. The lack of the ``small'' c hole in the Hartree--Fock (HF) description correcting the x hole is the main reason for the bad performance of HF for the dissociation of molecules.

It is instructive to consider the LDA holes, since they explain why DFT and its predecessor, X$\alpha$~\cite{Slater1951}, are so successful. The LDA holes are actually the $\lambda$-integrated ones, so a direct comparison is strictly not correct. Luckily, however, in the dissociation limit the $\lambda$-integrated and the exact hole at $\lambda = 1$ are identical~\cite{LeeuwenGritsenkoBaerends1996, PerdewSavinBurke1995}. The reason is that at long bond distances the interaction term $\lambda/r_{12}$ for $\lambda > 0$ will favor wavefunction configurations in which the electrons are residing on different atoms, i.e.\ the Heitler--London wavefunction. The density corresponding to this ground state wavefunction $\Psi_{\lambda}$ will be exactly equal to the wavefunction $\Psi_1$ at full coupling strength. It thus follows that $\Psi_{\lambda} = \Psi$ for $\lambda >0$ at infinite bond distance. At $\lambda = 0$ the wavefunction simply remains a KS determinant with a doubly occupied $\sigma_g$ orbital. However, the $\lambda=0$ region becomes unimportant~\cite{TealeCorianiHelgaker2009} in the $\lambda$-integration for the pair-correlation function and hence $g = \bar{g}$. Therefore at the longer bond distance of $R_{\text{H--H}} = 5.0$ Bohr the exact xc hole should be close to its $\lambda$-integrated counterpart.

In Fig.~\ref{fig:LDAholes5} we show the LDA holes for the hydrogen molecule at $R_{\text{H--H}} = 5.0$ Bohr. From our discussion in Sec.~\ref{sec:SymAndAsymp} it is clear what the definition of the xc hole of the LDA should be. The following expression~\cite{GunnarssonJonsonLundqvist1979, DreizlerGross1990} is consistent with the LDA energy expression~\eqref{eq:LDAenergy}
\begin{align}\label{eq:LDAhole}
\bar{\rho}^{\text{LDA}}_{\text{xc}}(\vecr|\vecr_{\text{ref}})
= n(\vecr_{\text{ref}})
\bigl(\bar{g}_h(n(\vecr_{\text{ref}}),\abs{\vecr - \vecr_{\text{ref}}}) - 1\bigr).
\end{align}
Gori--Giorgi and Perdew made an accurate model for the pair-correlation function of the homogeneous electron gas~\cite{Gori-GiorgiPerdew2002}, which we used to calculate $\bar{\rho}^{\text{LDA}}_{\text{xc}}$. The x hole can be calculated by using the exchange part of the electron gas pair-correlation function in this expression and the c hole is simply defined as the difference between the other two. The most striking feature of the LDA holes in Fig.~\ref{fig:LDAholes5} is that the x hole is localized instead of delocalized over the two atoms, just as is the case for the exact x hole. Although the LDA x hole does not resemble the exact x hole at all, the full xc hole (the one of interest) is actually modeled quite well. Especially, if we consider the exact x hole which is the hole employed in HF theory, the LDA xc hole provides a large improvement. We also see that the LDA c hole only provides a small correction to the x hole: it removes the outward oscillations and localizes the LDA hole a bit further. Since the correction from the c hole is so small, we can understand why the old X$\alpha$ method already outperformed HF so much. In particular, the deepening of the hole was empirically taken into account by scaling the exchange prefactor, $\alpha$, from its theoretical value, $2/3$, to $0.7$. These observations concerning the LDA holes also make it clear that it does not make sense at all to add ``exact exchange'' to LDA correlation. The same holds for GGA holes, since they are quite similar to LDA holes, only having slightly more wild oscillations and a discontinuity due to their cut-off to satisfy the sum-rule~\eqref{eq:xc holeSumRule} as additional features~\cite{SlametSahni1991, BurkePerdewWang1998}.

\section{The screened exchange Model}
\label{sec:SXmodel}

Considering these facts about the x hole in dissociating H$_2$, it seems to be unwise to add a correlation hole to an exact exchange hole. It will be hard to build a model for the correlation hole with the proper strongly varying features. But we also know that exchange effects often dominate and that correlation effects only provide a modification of the former. This observation suggests that we should not add correlation to exchange but rather modify the shape of the x hole by some correlation factor. From the holes for the hydrogen molecule (Figs~\ref{fig:H2holesEqui} and~\ref{fig:H2holes5}) we observe that the main effect of correlation is to \emph{localize} the x hole. This is not special for the H$_2$ molecule, but is the main feature of correlation in any system, even in the electron gas where the correlation reduces the long range behavior of the x hole from $r^{-4}$ to $r^{-8}$~\cite{Gori-GiorgiSacchettiBachelet2000a, Gori-GiorgiSacchettiBachelet2000b}. A further example was provided long ago by J.C. Slater when he pointed out that atomic term energies were often accurately described by term dependent Hartree--Fock theory (``exact exchange '') by simply reducing Slater's $F_k$ and $G_k$ integrals by $\sim25\%$~\cite{Slater1960}.

Following the discussion above it seems rather natural to use the following Ansatz for the xc energy $E_{\text{xc}}$ of an inhomogeneous system
\begin{align}\label{eq:ExcSX}
E_{\text{xc}} = -\frac{1}{4}\iinteg{\vecr}{\vecr'}\frac{\abs{\gamma_s(\vecr,\vecr')}^2}{\abs{\vecr-\vecr'}}F(\vecr,\vecr').
\end{align}
Here, $\gamma_s(\vecr,\vecr')$ is the spin-integrated non-interacting density matrix of the KS orbitals~\eqref{eq:KS1RDM} and the ``screening''-factor $F$ is intended to provide the necessary modification of exact exchange which will take care of the effects of correlations. The exact expression for $F$ is
\begin{align*}
F(\vecr,\vecr') \equiv \frac{\bar{\rho}_{\text{xc}}(\vecr|\vecr')}{\bar{\rho}_{\text{x}}(\vecr|\vecr')}
\end{align*}
and by multiplying the numerator and denominator by $n(\vecr)$ we immediately see that the screening function $F(\vecr,\vecr')$ is symmetric in its arguments $\vecr$ and $\vecr'$, cf.~\eqref{eq:xcHoleDef} and~\eqref{eq:xHoleDef}.

Our task is thus to find a reasonable model for $F$ and we stress again the importance of keeping the symmetry of $F(\vecr,\vecr')$ in order to have an ensuing xc potential with the correct $-1/r$ tail outside finite systems. We also believe it to be essential to have a model which satisfies the sum-rule for the xc hole and in terms of $F$, our model should thus obey
\begin{align*}
\integ{\vecr'}\abs{\gamma_s(\vecr,\vecr')}^2F(\vecr,\vecr') = 2n(\vecr),
\end{align*}
If we, like the founding fathers (KS), first turn to the electron gas, we realize that $F$, and also $\gamma_s$ for that matter, must be spherical functions of only $\abs{\vecr-\vecr'}$. In the spirit of the older WDAs we could thus attempt such an Ansatz for our $F$ function. It is, however, important here to stress that in the original WDAs it is almost the entire xc hole which is modeled in this way, whereas in our case we just model a modification of the full, in general non-spherical x hole. As mentioned previously, we would also like to model the $F$-function in the case of the dissociation of H$_2$. In the dissociation limit of the hydrogen molecule the $F$-function actually takes the form
\begin{align*}
F_{\text{HL}}(\vecr,\vecr') \approx f_a(\vecr)f_a(\vecr') + f_b(\vecr)f_b(\vecr')
\end{align*}
in terms of two one-point functions $f_a$ and $f_b$ located on the different  hydrogen atoms. This follows from the fact that in the limit of large separation between the nuclei, the Heitler--London wavefunction becomes exact---but we will not present the details here. But it means that in this limit, the $F$-function is very far from spherical.

When the two electrons are on different nuclei, the $F$-factor should vanish, because we are then dealing with two non-interacting subsystems and there is neither exchange nor correlation. When both electrons are on the same atom, the $F$-funcion should actually be 2 to make the xc hole equal to the negative of the local density. In this way the xc hole will precisely remove the full self-interaction on each atom---not just half the self-interaction as Hartree--Fock does---and we recover the correct limit of two separate and non-interacting atoms. 

As suggested by the above observations, the following Ansatz for the screening function $F(\vecr,\vecr')$ might stand a chance of carrying us from the limit of a homogeneous system to that of the complete breaking up of the H$_2$ bond
\begin{align}\label{eq:F_SX}
F^{\text{SX}}(\vecr,\vecr')
\equiv A(\vecr)A(\vecr')\, h\bigl(\abs{\vecr-\vecr'},\bar{n}(\vecr,\vecr')\bigr).
\end{align}
Here, the spherical function $h$ has an effective screening radius $\bar{r}_s$ given by $4\pi\bar{r}_s^3 = 3/\bar{n}$. This form was also inspired by the success of a wavefunction for the helium atom by Hylleraas~\cite{Hylleraas1929} having precisely this form.

We are then left with the choice of satisfying the hole sum-rule either by adjusting the one-point function $A(\vecr)$, the hole-depth function, or by varying the effective density $\bar{n}(\vecr,\vecr')$. We stress that the latter must be a symmetric two-point function in order not to jeopardize the asymptotics of the potential. In terms of the Ansatz~\eqref{eq:F_SX} the sum-rule reads
\begin{align}\label{eq:Asumrule}
A(\vecr)\integ{\vecr'}\abs{\gamma_s(\vecr,\vecr')}^2A(\vecr')\,
h\bigl(\abs{\vecr-\vecr'},\bar{n}(\vecr,\vecr')\bigr) = 2n(\vecr).
\end{align}
The effective density is expected to vary in a similar way as the density itself (from very small to very large values) and it is a two-point function. The hole-depth function $A$ on the other hand is a one-point function of limited variation---typically between zero and two depending on the normalization
of the function $h$. Consequently, it appears to be numerically more stable to use the $A$-function for the purpose of satisfying the sum-rule. Indeed, we have encountered no difficulties in solving~\eqref{eq:Asumrule} in our applications. A further argument in favor of this choice is a lack of guidance from the electron gas when determining the $A$-factor, should we have chosen to satisfy the sum-rule by varying the effective density $\bar{n}$. Most WDA models have used the latter procedure, but again, they have not considered an $A$-factor.

In our case we are thus free to choose the effective density. Typical choices are $\bar{n}(\vecr,\vecr') = \half\bigl(n(\vecr) + n(\vecr')\bigr)$ or $\bar{n}(\vecr,\vecr') = n\bigl((\vecr+\vecr')/2\bigr)$. in previous attempts to construct a symmetric version of an WDA-like model we have found that the first of these suggestions gave rise to numerical difficulties, whereas Gunnarsson et al.~\cite{GunnarssonJonsonLundqvist1979} encountered difficulties with the second choice. A choice which seems to work in our previous attempts is
\begin{align}\label{eq:avDensDef}
\bar{n}(\vecr,\vecr') \equiv \sqrt{n(\vecr)n(\vecr')},
\end{align}
which is the definition of the effective density $\bar{n}$ which we will use here. We stress, however, that there is no compelling reason for this choice. As a matter of fact, this is one part of our new model which we might have to revise in future attempts to improve the accuracy of the model.

It is not clear what properties and shape the screening function $h$ should have. However, since its main task is to screen the x hole, we will use the very simple form inspired by the screened Coulomb (Yukawa) interaction
\begin{align}\label{eq:HEGscreen}
h_{\text{HEG}}(r, \bar{n}) \equiv \e^{-D(\bar{n})r_{12}}.
\end{align}
The function $D(n)$ is fitted to the homogeneous electron gas such that it yields the exact xc energies for all homogeneous densities. In this way also the kinetic energy contribution to the xc energy is included. More details on the fit can be found in Appendix~\ref{ap:HEGscreen}. This form for the screening function is definitely too simplistic. More knowledge is required to build more accurate SX models. This will be the subject of future research.

\section{Illustrative results}
\label{sec:results}
One of the most severe tests for our new SX functional will be if it performs well for dissociating molecules. To keep matters simple, we have limited ourselves to the hydrogen molecule. The most natural grid for calculations on a diatomic molecule is based on a prolate spheroidal coordinate system. A planar elliptical grid with foci at the two nuclei and $z$-axis joining those foci is rotated about that axis to generate the full grid. More details on the grid used are in Appendix~\ref{ap:grid}. As a further simplification we used the density from a full CI  calculation in the same basis as used before (1s, 2s, 3s, 2p and 3d hydrogen wavefunctions). This CI expansion is not very good for obtaining an accurate total energy. However, we believe that the density will be accurate enough for the SX model.

\begin{figure}[t]
  \includegraphics[width=\columnwidth]{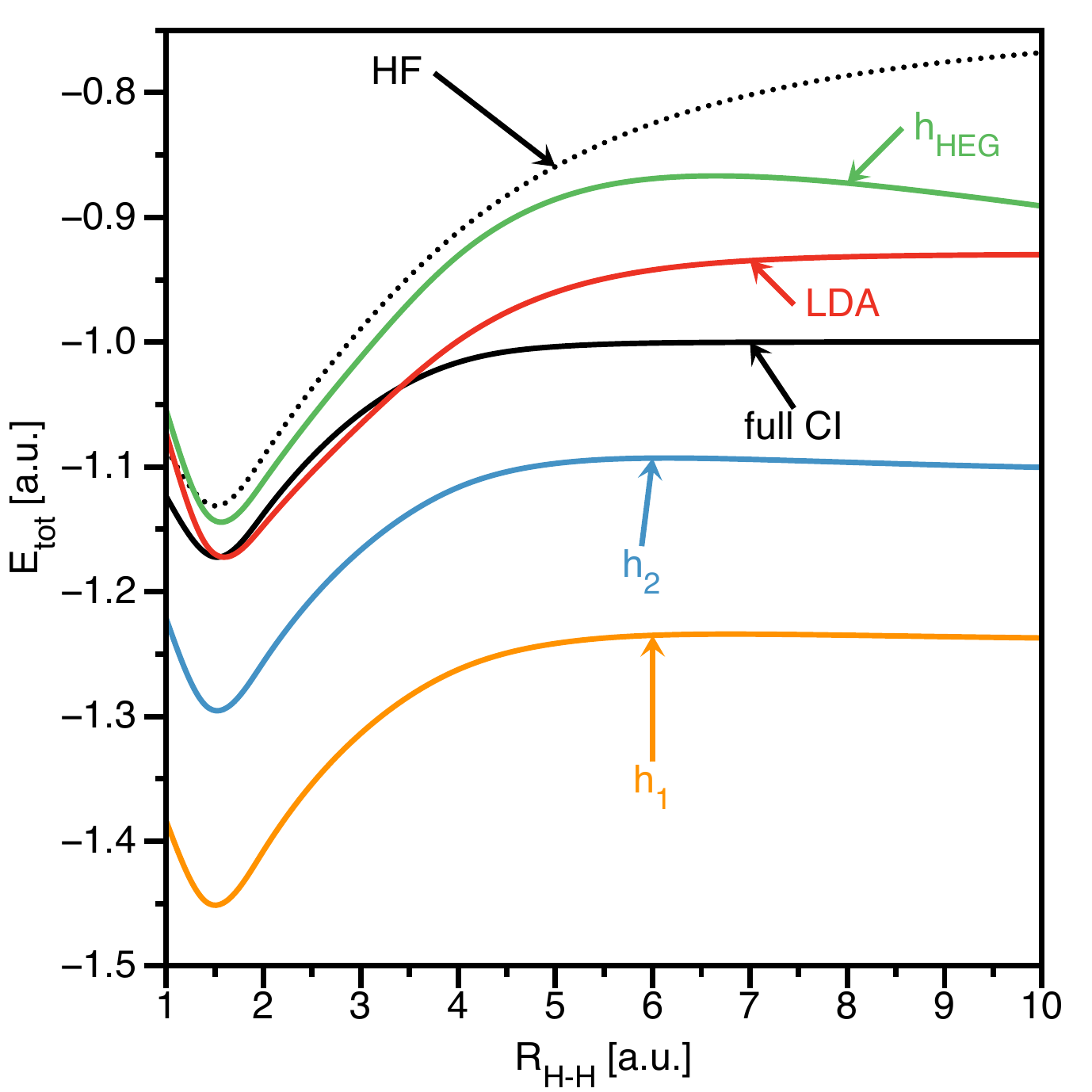} \\
  \caption{(Color online) Comparison of the total energy from the simple SX model with the ones from the full CI calculation for varying bond-length.}
  \label{fig:totalEnergies}
\end{figure}

\begin{figure*}[t]
  \includegraphics[width=\textwidth]{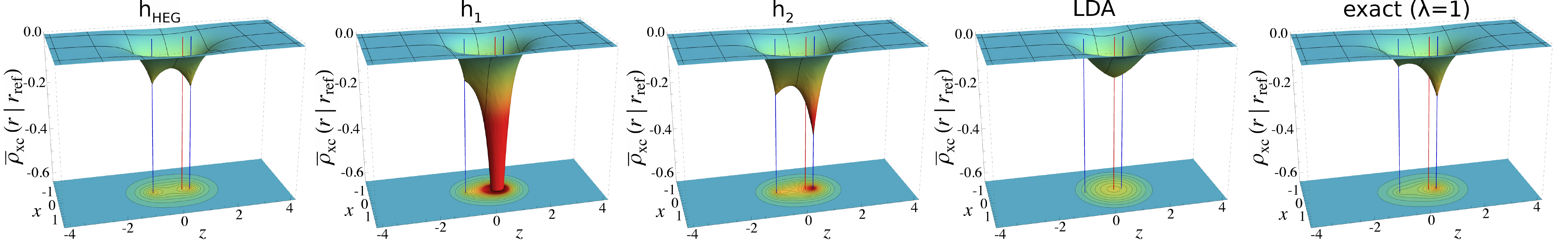} \\
  \caption{(Color online) The different model holes of the H$_2$ molecule at $R_{\text{H--H}} = 1.4$ Bohr in atomic units and the reference electron at 0.3 Bohr to the left of the right nucleus along the bond axis ($\rref = (0,0,0.4)$ Bohr). The positions of the nuclei are indicated by the blue lines and the position of the reference electron is indicated by the red line. The right most panel shows the exact (not integrated) xc hole for comparison. From left to right, the panels show the xc holes from the SX model with the $h_{\text{HEG}}$ screening function ~\eqref{eq:HEGscreen}, the SX model with the $h_1$ screening function~\eqref{eq:SlaterScreening}, the SX model with the $h_2$ screening function~\eqref{eq:GaussScreening} and the LDA hole evaluated as in~\eqref{eq:LDAhole} with the pair distribution from Ref.~\cite{Gori-GiorgiPerdew2002}.}
  \label{fig:modelHolesEqui}
  \includegraphics[width=\textwidth]{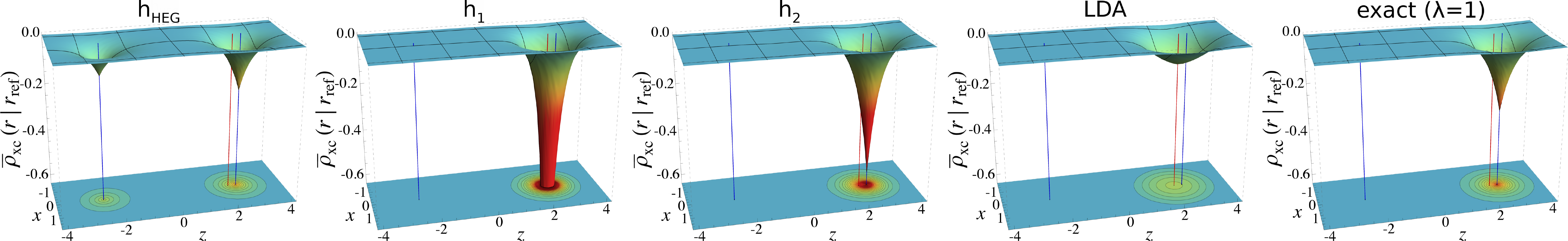}
  \caption{(Color online) Similar to the previous plots, but now for $R_{\text{H--H}} = 5.0$ Bohr. The reference electron is still at 0.3 Bohr to the left of the right nucleus along the bond axis ($\rref = (0,0,2.2)$ Bohr now).}
  \label{fig:modelHoles5}
\end{figure*}

The first step in evaluating the SX model is to solve for the hole-depth function, $A(\vecr)$, by solving the integral equation~\eqref{eq:Asumrule} on the grid. Once the hole-depth function is obtained, we performed the double integral~\eqref{eq:ExcSX} with $F^{\text{SX}}$~\eqref{eq:F_SX} to obtain the xc energy according to our simple SX model. To calculate the total energy, we note that the non-interacting kinetic energy can be directly obtained from the density for the two-electron system, $T_s = \half\int\bigl(\nabla\sqrt{n}\bigr)^2$, and the Hartree and nuclear contributions are already known from the full CI calculation. In Fig.~\ref{fig:totalEnergies} we compare the total energies from the SX model with the ones from a full CI calculation as a function of the distance, $R_{\text{H--H}}$, between the hydrogen atoms. As mentioned before the Slater basis is too poor to obtain a good total energy, so we performed a full CI calculation with the \textsc{dalton} package~\cite{Dalton} in an aug-cc-pVQZ basis~\cite{cc-pVT/QZ_H_B-Ne_aug_H} as a reference and additionally the corresponding HF result is shown as well. Unfortunately our SX model with the simplistic screening function is not performing much better than the HF method. It follows the HF curve rather closely and only around $R_{\text{H--H}} \approx 6$ Bohr does the curve start to bend downwards to the correct total energy. The most important feature of the SX model is that it directly models the xc hole and therefore we can look at it what is going wrong. In Fig.~\ref{fig:modelHolesEqui} in the left panel, we show the hole at $R_{\text{H--H}} = 1.4$ Bohr. If we compare with the exact holes in Fig.~\ref{fig:H2holesEqui}, we see immediately that our current SX model is not localizing the x hole sufficiently. Actually, the SX hole is almost identical to the x hole $\bar{\rho}_x$. In the left panel of Fig.~\ref{fig:modelHoles5} we show the hole the elongated distance $R_{\text{H--H}} = 5.0$ Bohr. Comparing with the exact holes in Fig.~\ref{fig:H2holes5} we see that the SX model actually does localize the x hole, but not sufficiently. The lack of localization of the hole explains exactly why the total energy in the SX model is consistently too high (Fig.~\ref{fig:totalEnergies}). Due to the $1/\abs{\vecr-\vecr'}$ term in the expression for the xc energy~\eqref{eq:xcEnergy}, a too delocalized hole does not stabilize the energy enough, which results in a too high energy.

In retrospect we should not be surprised that parametrizing the screening function by the homogeneous electron gas did not work out. The main task of the screening function is to localize the hole which is most important in inhomogeneous systems like the hydrogen molecule. Therefore, its actual form and localization strength should be modeled by these systems and not the homogeneous electron gas. A detailed study and parametrization are beyond the aims of this article, but to reinforce our arguments for this statement, we also like to show some results for the following two heuristic screening functions
\begin{subequations}
\begin{align}
\label{eq:SlaterScreening}
h_1(r_{12},\bar{n}) &= \exp\biggl(-c_1\Bigl(\frac{r_{12}}{\bar{r}_s}\Bigr)\biggr), \\
\label{eq:GaussScreening}
h_2(r_{12},\bar{n}) &= \exp\biggl(-c_2\Bigl(\frac{r_{12}}{\bar{r}_s}\Bigr)^2\biggr).
\end{align}
\end{subequations}
The first one, $h_1$, keeps the Slater like form, but replaces the parametrization by the electron gas by a simple division by $\bar{r}_s$ to make the total dimensionless and a constant $c_1$ that we can choose to our liking. We found that $c_1 = 2.0$ gave a nice dissociation behavior for the energy. The second one is mainly included to emphasize that the required shape of the screening function is also unclear at the moment. Choosing the constant in the same manner as before, we found $c_2 = 0.5$ to be sufficient for our purposes. Note that both screening functions are not fitted to the electron gas anymore and are therefore not expected to give the correct xc energy density, $\varepsilon_{\text{xc}}(r_s)$, for the gas.

\begin{figure}[t]
  \includegraphics[width=\columnwidth]{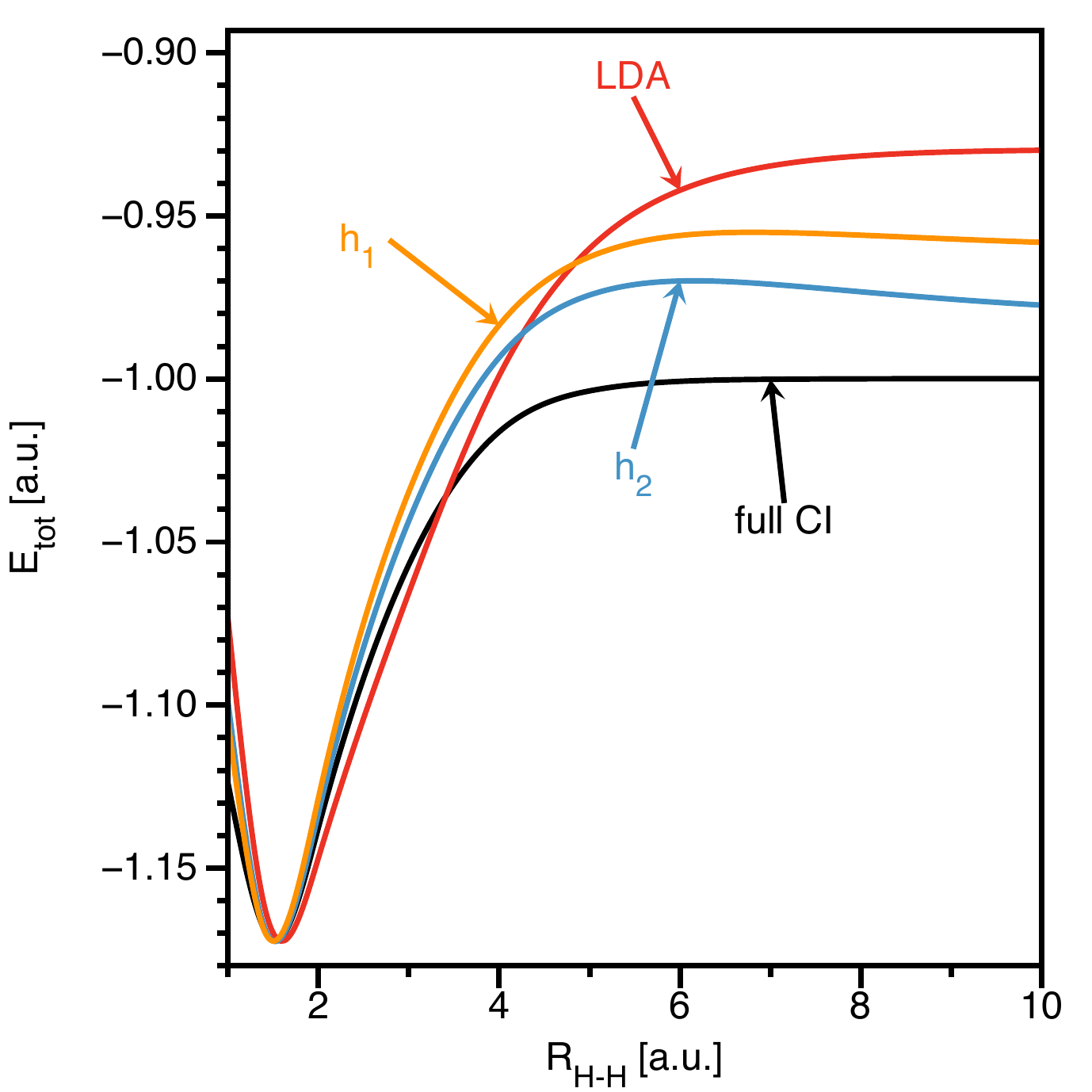} \\
  \caption{(Color online) Comparison of the total energy of the LDA functional~\cite{Gori-GiorgiPerdew2002} and the SX model with the Gaussian, $h_2$, as its screening function with the full CI results. The total energy of the LDA and SX model are shifted such that the minima coincide with the full CI value.}
  \label{fig:shiftedEnergies}
\end{figure}

The results for the energy from these screening functions are shown also in Fig.~\ref{fig:totalEnergies}. The situation is now rather different. The total energy is consistently underestimated. However, the shape of the curve is definitely an improvement. The total energy at elongated distances $R_{\text{H--H}} > 5$ Bohr remains rather constant as we choose the constants $c_i$ to do so. From the figure it is not immediately clear if the shape is also an improvement over the simple LDA functional. However, shifting the curves such that their minima coincide with the full CI minimum (Fig.~\ref{fig:shiftedEnergies}), we see that the energy from the SX models follow the full CI energy much closer. In particular the Gaussian, $h_2$ seems to reduce the overbinding most. Even at equilibrium distance the position and shape of the minimum seems to be somewhat improved.

Again, since we have built a direct model for the hole we can explain both features. Considering the holes from the heuristic $h_1$ and $h_2$ screening functions in Figs~\ref{fig:modelHolesEqui} and~\ref{fig:modelHoles5}, we see that both screening function are too powerful: they screen the x hole too much. This explains why the total energy is consistently lower than the full CI result: the hole becomes too compact. Since the hole is even more compact for $h_1$ screening function than the $h_2$ screening function, the energy of the $h_1$ screening function is even lower than the energy of the $h_2$ screening function. However, if we consider the shapes of the holes, then they are definitely an improvement over the previous screening function. Especially at the elongated bond distance $R_{\text{H--H}} = 5.0$ Bohr, we see that both heuristic screening functions nicely localize the hole completely at the correct side, which explains their improved energy in the dissociation limit over the erroneous $1/R_{\text{H--H}}$ of HF.

The LDA hole is also shown in Figs~\ref{fig:modelHolesEqui} and~\ref{fig:modelHoles5}. Compared to the HF, cf.\ $\bar{\rho}_x$ in Fig.~\ref{fig:H2holesEqui} and~\ref{fig:H2holes5}, the LDA hole is a big improvement since it localizes around the reference electron. This improvement is also visible in the total energy where the LDA does not exhibit the erroneous $1/R_{\text{H--H}}$ behavior as HF does (Fig.~\ref{fig:totalEnergies}). However, the shape of the LDA hole is ridiculous. It is spherical by construction and it does not have the peaked features at the nuclei. Instead, the LDA hole attains its minimum at the position of the reference electron. The SX holes, especially with the $h_2$ screening function, are an improvement over both the HF hole and the LDA hole. It correctly localizes the hole around the reference electron and still retains the distinct peaked features of the hole. Hence, the binding curve is much improved over HF and LDA (see Fig.~\ref{fig:shiftedEnergies}).

\section{Conclusion}
\label{sec:conclusion}
The aim of the present work is to construct a functional for the exchange-correlation (xc) energy of DFT which applies to such diverse systems as the electron gas and the dissociation of the hydrogen molecule (H$_2$). To this end we try to find a model for the xc hole of these and intermediate systems in real space. It has been known for long that the exchange (x) hole captures many important features of the full one. For instance, in atoms term energies are often well described by reducing the exchange effects by $25\%$ and in the electron gas correlations have the effect of reducing the range of the pure x hole from a $r^{-4}$ decay to a $r^{-8}$ decay~\cite{Gori-GiorgiSacchettiBachelet2000a, Gori-GiorgiSacchettiBachelet2000b}, $r$ being the distance from the the center of the hole. Thus unlike previous models that have sought to model the xc hole in real space, our present model aims at modifying the full x hole. Our investigations have shown that in the case of the dissociation of H$_2$ the xc hole is qualitatively different from the x hole. Therefore, it is an unwise strategy to add a correlation (c) hole to a full x hole. The former would have to replace the latter with a full xc hole with appropriate features. We show here that this can be achieved in a more natural manner by multiplying the x hole by an appropriate correlation factor. Judging from the electron gas, the correlation factor might be chosen as a function of the distance $\abs{\vecr-\rref}$ between an electron and a reference electron. (We here again remind the reader that the xc hole describes the depletion of negative charge around an electron  known to be at the reference position $\rref$.) Unfortunately, our investigations have shown that such a simple correlation factor will have difficulties in moving half an x hole from one atom to the other---which is the appropriate effect of correlation in H$_2$ at large nuclear separation. Instead, we have chosen to include in our correlation factor an additional factor $A(\vecr)$ depending on only one coordinate---a modulation of the depth of the xc hole. Such a factor is by symmetry just a constant in the electron gas and irrelevant to the shape of the hole in that case. Consequently, we have no guidance from the gas in choosing the $A$-factor. Instead, we have decided to determine this factor at every point in space from the sum-rule for the xc hole. This sum-rule is of course of utmost importance for obtaining an xc potential with the correct asymptotics outside finite systems ($-1/r$) from our model. This important property also requires the full symmetry in $\vecr$ and $\vecr'$ in the density multiplied xc hole, $n(\vecr)\bar{\rho}_{\text{xc}}(\vecr'|\vecr)$, something that we have emphasized throughout the paper.

The decision to use the $A$-factor for satisfying the sum-rule leaves us with great freedom in choosing a screening factor depending only on $\abs{\vecr-\vecr'}$. Thinking about the electron gas it is natural to let this screening function have a screening length $\bar{r}_s$ determined by an effective density $\bar{n}$ according to $4\pi\bar{n}\bar{r}_s^3 = 3$. In the present work we have made the somewhat arbitrary choice $\bar{n}(\vecr,\vecr') = \sqrt{n(\vecr)n(\vecr')}$. We are, however, aware of that we might be forced to abandon this simple choice in future refinements of our model.

For the actual shape of the screening function we have simply made a couple of reasonable choices for illustrational purposes. They are rapidly decaying, analytic functions with a few parameters with density dependence. One of the screening functions had its parameters specifically chosen such to reproduce the ``exact'' xc energies of the electron gas in the homogeneous limit, whereas two other screening functions had more \emph{ad-hoc} parameters to illustrate the effect of selecting different forms of the screening function.

We could, however, also have attempted to find a screening function with a shape that would have allowed us to obtain accurate results for one- and two-electron systems. We would then have had an approximation which is able to properly dissociate H$_2$, which would be exact for the electron gas and quite accurate for the one- and two-electron cases. Such a functional is likely to give good total energies in a large number of systems.

In the present work we have, however, refrained from going down this road. Instead, our aim here was to present the basic ideas and to leave further refinements to future investigations. In order to just illustrate our new approach, we thus chose to present results obtained from two simple, but rather arbitrary screening functions given by the equations~\eqref{eq:SlaterScreening} and~\eqref{eq:GaussScreening}. We have seen that the shorter ranged choice ($h_2$) gives better results, but they are still not very good. It is however, seen from both Fig.~\ref{fig:totalEnergies} and Fig.~\ref{fig:shiftedEnergies} that the errors are mainly associated with an inadequate description of a simple one-electron system. (At a bond distance of 10 Bohr we basically have two separate hydrogen atoms.)

The most important result of the present work is that we managed to design a model which is able to describe the proper behavior of the xc hole of a hydrogen molecule as it dissociates. The details are not overly accurate, but we nurture real hope that they may fall in place by a better choice of the screening function. For now, however, we have left the search for such a function to future work.

\appendix

\section{The xc potential}
\label{ap:vxc}
The xc potential is obtained by straightforward differentiation of $E^{\text{SX}}_{\text{xc}}$ with respect to the density
\begin{multline}\label{eq:vSXxc}
v^{\text{SX}}_{\text{xc}}(\vecr)
= -\frac{1}{4}\iinteg{\vecr_1}{\vecr_2}\frac{1}{\abs{\vecr_1-\vecr_2}}  \\
{}\times\biggl(\frac{\delta\abs{\gamma_s(\vecr_1,\vecr_2)}^2}{\delta n(\vecr)}
A(\vecr_1)A(\vecr_2)h(r_{12},\bar{n})  \\
{} + 2\abs{\gamma_s(\vecr_1,\vecr_2)}^2\frac{\delta A(\vecr_1)}{\delta n(\vecr)}
A(\vecr_2)h(r_{12},\bar{n}) \\
{} + \abs{\gamma_s(\vecr_1,\vecr_2)}^2A(\vecr_1)A(\vecr_2)
\frac{\delta h(r_{12},\bar{n})}{\delta n(\vecr)}\biggr).
\end{multline}
The functional derivative of the hole-depth function can be obtained by differentiating its hole sum-rule~\eqref{eq:Asumrule} with respect to the density. Its derivative can be worked out as
\begin{multline*}
\frac{\delta A(\vecr_1)}{\delta n(\vecr)} +
\frac{A^2(\vecr_1)}{2n(\vecr_1)}\integ{\vecr_2}\abs{\gamma_s(\vecr_1,\vecr_2)}^2h(r_{12},\bar{n})
\frac{\delta A(\vecr_2)}{\delta n(\vecr)} \\
= \frac{A(\vecr)}{n(\vecr)}\delta(\vecr-\vecr_1) -
\frac{A^2(\vecr_1)}{2n(\vecr_1)}\integ{\vecr_2}A(\vecr_2) \times {} \\
\biggl(\frac{\abs{\gamma_s(\vecr_1,\vecr_2)}^2}{\delta n(\vecr)}h(r_{12},\bar{n}) + 
\abs{\gamma_s(\vecr_1,\vecr_2)}^2\frac{\delta h(r_{12},\bar{n})}{\delta n(\vecr)}\biggr).
\end{multline*}
Unfortunately this equation does not give a closed expression for the functional derivative of $A(\vecr)$, but just like its original counterpart~\eqref{eq:Asumrule} has to be solved iteratively to self-consistency.

Since the hole correctly integrates to $-1$ electron~\eqref{eq:Asumrule}, the potential has the correct asymptotic behavior $-1/r$. However, due to all the functional derivatives this is not directly visible in the expression for the xc potential~\eqref{eq:vSXxc}. The main complication arises from the functional derivative of the KS density matrix whose evaluation requires the application of the chain-rule multiple times. Fortunately, in the case of a two-electron system, the Kohn--Sham density matrix can be expressed directly in the density as $\gamma_s(\vecr_1,\vecr_2) = \sqrt{n(\vecr_1)n(\vecr_2)}$ allowing for direct differentiation. Working out the first part of the potential gives
\begin{align}
v^{\text{SX}}_{\text{xc}}(\vecr)
= -\frac{A(\vecr)}{2}\integ{\vecr'}\frac{n(\vecr')A(\vecr_2)h(r_{12},\bar{n})}{\abs{\vecr-\vecr'}} + \dotsb,
\end{align}
so we find the correct Coulombic $1/r$ behavior. We only have to check the charge. Using $\gamma_s(\vecr_1,\vecr_2) = \sqrt{n(\vecr_1)n(\vecr_2)}$ in the sum-rule for the hole-depth function~\eqref{eq:Asumrule}, we find
\begin{align}
\frac{A(\vecr)}{2}\integ{\vecr'}n(\vecr')A(\vecr')h\bigl(\abs{\vecr-\vecr'},\bar{n}\bigr) = 1,
\end{align}
so indeed, the Coulombic part of the potential also has the correct charge of $-1$.

\section{The electron gas screening function}
\label{ap:HEGscreen}
One of the requirements of the new functional is that it should work form a broad class of systems. Not only for molecules, but also for extended systems and in particular the homogeneous electron gas. Since the homogeneous electron gas is well studied and much of its properties are known, it provides the ideal system to fit the screening function~\eqref{eq:HEGscreen} such that the SX functional will be exact for the homogeneous electron gas, in particular, it should give the exact xc energy for the electron gas. Before we can start fitting the screening function, we have to solve for the hole-depth, $A$. For the KS density matrix of the homogeneous electron gas one can straightforwardly calculate that
\begin{align*}
\gamma_s(r_{12},k_F)
= \frac{k_F^3}{\pi^2}\frac{\sin(k_Fr_{12}) - k_Fr_{12}\cos(k_Fr_{12})}{(k_Fr_{12})^3},
\end{align*}
where the Fermi wavevector $k_F^3 \equiv 3\pi^2n$.
Using this expression for the KS density matrix in the sum-rule for the hole-depth function~\eqref{eq:Asumrule} and the Ansatz for the screening function~\eqref{eq:HEGscreen}, we find
\begin{align}\label{eq:AelGas}
1 = \frac{6}{\pi} A^2 F_4(\tilde{D}),
\end{align}
where we defined $\tilde{D} \equiv D / k_F$ and the integrals
\begin{align*}
F_n(\beta) \equiv \int_0^{\infty}\!\!\!\ud y \, \bigl(\sin(y) - y\cos(y)\bigr)^2y^{-n}\e^{-\beta y}.
\end{align*}
More details about these functions and explicit expressions are given in Appendix~\ref{ap:Fn_functions}. Now we have an explicit relation how to calculate the hole-depth function $A$ for the homogeneous electron gas~\eqref{eq:AelGas}, we will fix $D$ by requiring our functional to produce the exact xc energy for the electron gas. In terms of our model, the xc energy density becomes
\begin{align}\label{eq:DtildeExcHEG}
\varepsilon_{\text{xc}}
= -\frac{3k_F}{\pi}A^2F_5(\tilde{D}).
\end{align}
Since the equations~\eqref{eq:AelGas} and~\eqref{eq:DtildeExcHEG} are linear in $A^2$, the hole-depth function can simply be eliminated by dividing the equations, which gives
\begin{align}\label{eq:eps_xcCond}
\frac{F_5(\tilde{D})}{F_4(\tilde{D})} 
= \frac{3}{2\pi}\frac{\varepsilon_{\text{xc}}(r_s)}{\varepsilon_{\text{x}}(r_s)},
\end{align}
where we used the Seitz radius $r_s^3 \equiv 3/(4\pi n)$ and that $\varepsilon_{\text{x}} = -3k_F/(4\pi)$ for the homogeneous electron gas. To find $\tilde{D}(r_s)$, we need an explicit expression for $\varepsilon_{\text{xc}}(r_s)$ for which we used an accurate fit to the random-phase approximation (RPA) and Green's function Monte Carlo data by Perdew and Wang~\cite{PerdewWang1992a}. Unfortunately, the expression for $\tilde{D}(r_s)$ cannot be inverted analytically. However, one can show that $F_5(\beta)/F_4(\beta)$ is a monotonic increasing function over $\beta \geq 0$, so at least the solution to~\eqref{eq:eps_xcCond} is unique. Further, in the low density limit we have
\begin{align*}
\lim_{r_s \to \infty} \frac{3}{2\pi}\frac{\varepsilon_{\text{xc}}(r_s)}{\varepsilon_{\text{x}}(r_s)}
= \frac{3}{2\pi}\left(1+\frac{4\pi}{3}\sqrt[3]{\frac{4}{9\pi}}\frac{\alpha_1}{\beta_4}\right) = 0.92925,
\end{align*}
where the parameters $\alpha_1 = 0.21370$ and $\beta_4 = 0.49294$ are from the low-density fit of Perdew and Wang~\cite{PerdewWang1992a}. Inverting~\eqref{eq:eps_xcCond} numerically, we find that in the low density limit
\begin{align*}
D(r_s) \approx \frac{D^{\infty}}{r_s},
\end{align*}
where $D^{\infty} = 2.27591$. For the asymptotic behavior in the high density limit, $r_s \to 0$, we find for the ratio of the xc and x energies
\begin{align*}
\frac{3}{2\pi}\frac{\varepsilon_{\text{xc}}(r_s)}{\varepsilon_{\text{x}}(r_s)} 
= \frac{3}{2\pi} - 2\sqrt[3]{\frac{4}{9\pi}}c_0\,r_s\ln(r_s) + \dotsb,
\end{align*}
where $c_0 = c_0^{\text{RPA}} = \bigl(1 - \ln(2)\bigr) / \pi^2$ is a constant from the high-density fit by Perdew and Wang to the RPA~\cite{PerdewWang1992a}. Similarly for small arguments of $F_5/F_4$ we have
\begin{align*}
\frac{F_5(\tilde{D})}{F_4(\tilde{D})}
= \frac{3}{2\pi} - \frac{9}{2\pi^2}\sqrt[3]{\frac{4}{9\pi}}D(r_s)\,r_s\ln(r_s) + \dotsb.
\end{align*}
Comparing these two high density limits, we find that for high density to lowest order
\begin{align*}
D(r_s) \approx D^0 \equiv \frac{4}{9}\bigl(1 - \ln(2)\bigr).
\end{align*}
To obtain a workable expression for $D(r_s)$, we simply solved~\eqref{eq:eps_xcCond} numerically for several $r_s$ and made a least square fit using the following the following Padé approximant
\begin{align*}
D(r_s) = \frac{a_0 + a_1r_s + b_3D^{\infty}r_s^2}{1 + b_1r_s + b_2r_s^2 + b_3r_s^3},
\end{align*}
for which we found the coefficients $a_0 = 0.149056$, $a_1 = 0.180374$, $b_1 = 1.16435$, $b_2 = 0.128538$ and $b_3 = 0.000703698$. Note that we did not enforce the proper limit for $r_s \to 0$, since the low density region is more important and relaxing this constraint significantly increased the accuracy for $r_s > 0.1$ Bohr, which are more relevant densities in non-relativistic molecules and solids.

\section{The prolate spheroidal grid}
\label{ap:grid}
In this appendix we very briefly introduce the prolate spheroidal grid and give details on the exact parameters used for the grid in the calculations. The prolate spheroidal grid is created from an elliptic grid by rotating it around the axis connecting the two foci, the $z$-axis. The elliptic coordinates are defined as
\begin{align*}
\xi &\equiv \frac{r_1 + r_2}{2\rho} &
&\text{and} &
\eta &\equiv \frac{r_1 - r_2}{2\rho},
\end{align*}
where $r_i$ is the distance to nucleus $i$, which are placed at $\pm\rho$ from the origin on the $z$-axis. One readily sees that the ranges of the elliptic coordinates are $1 \leq \xi$ and $-1 \leq \eta \leq 1$. The curves $\xi = \text{constant}$ describe ellipses and the $\eta = \text{constant}$ curves describe hyperbolae. The intersection points of the ellipses and hyperbolae for different constants will be used as grid points. Also taking the revolution around the $z$-axis into account one can derive expressions for the cartesian coordinates in therms of the prolate spheroidal ones
\begin{align*}
x	&= \rho\sqrt{(\xi^2-1)(1-\eta^2)}\cos(\phi), \notag \\
y	&= \rho\sqrt{(\xi^2-1)(1-\eta^2)}\sin(\phi), \\
z	&= \rho\xi\eta. \notag
\end{align*}
The main advantage in using ellipses and hyperbolae is that they are orthogonal to each other. Therefore, the metric $\mat{g}$ is simply diagonal, $g_{\xi\eta} = g_{\eta\phi} = g_{\phi\xi} = 0$, with diagonal elements
\begin{align*}
g_{\xi\xi} &= \abraket{\frac{\du\vecr}{\du\xi}}{\frac{\du\vecr}{\du\xi}} 
= \rho^2\frac{\xi^2-\eta^2}{\xi^2-1}, \notag \\
g_{\eta\eta} &= \abraket{\frac{\du\vecr}{\du\eta}}{\frac{\du\vecr}{\du\eta}} 
= \rho^2\frac{\xi^2-\eta^2}{1-\eta^2}, \\
g_{\phi\phi} &= \abraket{\frac{\du\vecr}{\du\phi}}{\frac{\du\vecr}{\du\phi}} 
= \rho^2(\xi^2-1)(1-\eta^2). \notag
\end{align*}
The scaling factors are defined as $\lambda_i \equiv \sqrt{g_{ii}}$, from which we immediately find the volume element
\begin{align*}
\Omega = \lambda_{\xi}\lambda_{\eta}\lambda_{\phi} = \rho^3(\xi^2-\eta^2)
\end{align*}
and the gradient
\begin{align*}
\nabla &= \begin{pmatrix}
\dfrac{1}{\lambda_{\xi}}\dfrac{\du}{\du\xi} \\
\dfrac{1}{\lambda_{\eta}}\dfrac{\du}{\du\eta} \\
\dfrac{1}{\lambda_{\phi}}\dfrac{\du}{\du\phi}
\end{pmatrix}
= \frac{1}{\rho}\begin{pmatrix}
\sqrt{\dfrac{\xi^2-1}{\xi^2-\eta^2}}\dfrac{\du}{\du\xi} \\
\sqrt{\dfrac{1-\eta^2}{\xi^2-\eta^2}}\dfrac{\du}{\du\eta} \\
\dfrac{1}{\sqrt{(\xi^2-1)(1-\eta^2)}}\dfrac{\du}{\du\phi}
\end{pmatrix}.
\end{align*}
We could also write down an explicit expression for the Laplacian, but we do not need it. The disadvantage of discretizing the Laplacian directly is that it is not symmetric anymore. Instead we use that for functions vanishing sufficiently fast at the boundary~\cite{Becke1982}
\begin{align*}
-\integ{\vecr} f(\vecr)\nabla^2g(\vecr)
= \integ{\vecr} \nabla f(\vecr)\cdot \nabla g(\vecr),
\end{align*}
so formally we can write
\begin{align*}
-\Omega(\vecr)\nabla^2
= \leftnabla\sqrt{\Omega(\vecr)}\cdot\sqrt{\Omega(\vecr)}\,\rightnabla,
\end{align*}
which is inherently symmetric and its discretization can directly be constructed from the discretized gradient.

The $\xi$-grid points are generated by starting from an equidistant grid, $u \in [0,1)$, and using a simplified version of the transformation used by Becke~\cite{Becke1982} to generate grid points and weights for the $\xi$-grid
\begin{align*}
\xi(u) = \frac{1}{(1-u^2)^{p_{\xi}}}.
\end{align*}
The parameter $p_{\xi}$ can be used to affect the distribution of the grid points between the inner and outer region. For a given maximum value of $\xi$, $\xi_{\text{max}}$, $p_{\xi}$ is easily determined as
\begin{align*}
p_{\xi} = -\frac{\ln(\xi_{\text{max}})}{\ln\bigl(\Delta u(2 - \Delta u)\bigr)},
\end{align*}
where $\Delta u$ is the distance between the points in the $u$-grid. An important aspect is the $u^2$ which has the benefit that the cusps at the nuclei are transformed into smooth Gaussians in $u$-grid. The disadvantage is that the weight becomes zero at the line between the nuclei ($w(\xi = 1) = \xi'(u=0) = 0$), so the solution cannot be calculated directly at these points and has to be extrapolated.

For the $\eta$-grid we started from an equidistant grid, $v \in [0, \pi]$, and use the following transformation to generate the $\eta$-grid points
\begin{align*}
\eta(v) &= -\cos\bigl(v + p_{\eta}\sin(2v)\bigr),
\end{align*}
where the parameter $p_{\eta}$ can be used to modify the point distribution between the intermediate region and the nuclei. Becke found $p_{\eta} = -0.25$ to be optimal~\cite{Becke1982}. Also this transformation has the property that the cusps at the  nuclei are transformed into Gaussians in the $v$-grid. As one might already expect, the disadvantage of this feature is that the weights at the boundaries vanish ($w(\eta=-1) = w(\eta=1) = 0$), so the results will have to be extrapolated also to these points. The $\eta=\pm1$ points correspond along the bond axis outside the molecule.

We found that 80, 81 and 40 points in the $\xi$, $\eta$ and $\phi$-directions respectively gave numerically sufficient converged results. For the self-consistent solution of the hole-depth equation from its sum-rule~\eqref{eq:Asumrule} it was important that the density did not become too small. Hence the grid should not have points far in the asymptotic region. Using $\xi_{\text{max}} = 1 + 10/\rho$ was sufficient to prevent points with numerically vanishing density, though still including a sufficient part of the relevant space.

In the equidistant grids it is straightforward to define numerical derivatives. For example one can use cubic B-splines~\cite{Becke1982} or simple central finite differences~\cite{Davstad1990} which we used in our calculations. In practical calculations, 4th order derivatives already gave sufficient accuracy.

\section{The functions $F_n$}
\label{ap:Fn_functions}
In this appendix we show how the integrals defining the functions $F_n$ can be evaluated. The functions $F_n(\beta)$ satisfy
\begin{align}
\frac{\ud F_{n+1}(\beta)}{\ud \beta} = -F_n(\beta).
\end{align}
Using the boundary condition $F_n(+\infty) = 0$, we find
\begin{align}\label{eq:FnIntegral}
F_{n+1}(\beta) = \int_{\beta}^{\infty}\!\!\!\ud u \, F_n(u),
\end{align}
which can be used to obtain successively higher order functions. The integral most suitable for direct evaluation is $F_0$, albeit the evaluation is still rather tedious. The final result is
\begin{align}
F_0(\beta) = 16 \frac{5\beta^2 + 4}{\beta^3(\beta^2+4)^3}.
\end{align}
By applying successively the integral formula~\eqref{eq:FnIntegral} we can obtain the higher order integrals required for our model. Their evaluation is straightforward, but takes the necessary amount of time. The final results are
\begin{widetext}
\begin{align*}
F_1(\beta) &= \frac{\ln(1+4/\beta^2)}{4} + \frac{1}{2\beta^2} - \frac{4}{(\beta^2+4)^2} - \frac{3/2}{\beta^2+4}, \\
F_2(\beta) &= \frac{1}{4}\arctan(\beta/2) + \half\frac{\beta}{\beta^2+4}, \\
F_3(\beta) &= \frac{(\beta^2 + 2)\ln(1+4/\beta^2)}{8} - \half, \\
F_4(\beta) &= \frac{\arctan(2/\beta)}{3} - \frac{(\beta^3 + 6\beta)\ln(1+4/\beta^2)}{24} +
\frac{\beta}{6}, \\
F_5(\beta) &= \frac{1}{4} + \frac{\beta^2(\beta^2+12)\ln(1+4/\beta^2)}{96} - \frac{\beta^2}{24} - 
\frac{\beta\arctan(2/\beta)}{3}, \\
F_6(\beta) &= \frac{(5\beta^2+4)\arctan(2/\beta)}{30} - \frac{\beta^3(\beta^2 +20)\ln(1+4/\beta^2)}{480} + \frac{\beta^3}{120} - \frac{11\beta}{60}.
\end{align*}
\end{widetext}
The integrals for $n > 6$ do not converge anymore, so $F_6(\beta)$ is actually the last function in this series.

\begin{acknowledgments}
K.J.H.G. would like to thank N.E. Dahlen for writing the code to calculate the Slater integrals needed in the full CI program. K.J.H.G. and R.v.L. acknowledge the Academy of Finland for research funding under Grant No.\ 127739.
\end{acknowledgments}

\bibliography{bible}

\end{document}